\documentclass[a4paper,fleqn,usenatbib]{mnras}
\usepackage{newtxtext,newtxmath}
\usepackage[T1]{fontenc}
\usepackage{ae,aecompl}
\usepackage{graphicx}	
\usepackage{amsmath}	

\newcommand{\beq}{\begin{equation}}
\newcommand{\eeq}{\end{equation}}

\def\asec{\ifmmode ^{\prime\prime}\else$^{\prime\prime}$\fi}

\def\msun{\hbox{M$_{\odot}$}}

\def\msunyr{\mbox{\,${\rm M_{\odot}\, yr^{-1}}$}}
\def\mdot{\dot M}

\def\degs{\ifmmode ^{\circ}\else$^{\circ}$\fi}
\def\amin{\ifmmode ^{\prime}\else$^{\prime}$\fi}
\def\asec{\ifmmode ^{\prime\prime}\else$^{\prime\prime}$\fi}

\def\degs{\ifmmode ^{\circ}\else$^{\circ}$\fi}
\def\amin{\ifmmode ^{\prime}\else$^{\prime}$\fi}

\def\cm{\mbox{\,cm}}
\def\ccc{\mbox{\,cm$^{-3}$}}
\def\cm3{\mbox{\,cm$^{-3}$}}
\def\kms{\mbox{\,km~s$^{-1}$}}

\def\kms{\mbox{\,km s$^{-1}$}}

\def\pccc{\mbox{\,pc cm$^{-3}$}}
\def\lsim{\!\!\!\phantom{\le}\smash{\buildrel{}\over
 {\lower2.5dd\hbox{$\buildrel{\lower2dd\hbox{$\displaystyle<$}}\over
                                 \sim$}}}\,\,}
\def\gsim{\!\!\!\phantom{\ge}\smash{\buildrel{}\over
{\lower2.5dd\hbox{$\buildrel{\lower2dd\hbox{$\displaystyle>$}}\over
                               \sim$}}}\,\,}

\def\aj{AJ}                   
\def\araa{ARA\&A}             
\def\apj{ApJ}                 
\def\apjl{ApJ}                
\def\apjs{ApJS}               
\def\aap{A\&A}                
\def\aapr{A\&A~Rev.}          
\def\mnras{MNRAS}             
\def\nat{Nature}              
\def\pasp{PASP}               

\title[Is FRB 191001 embedded in a supernova remnant? ]{Is FRB 191001 embedded in a supernova remnant?}
\author[]{
Esha Kundu$^{1,2}$\thanks{kunduesh@msu.edu}
\\
$^{1}$International Centre for Radio Astronomy Research, Curtin University, Bentley, WA 6102, Australia\\
$^{2}$Center for Data Intensive and Time Domain Astronomy, Department of Physics and Astronomy, Michigan State University, East Lansing, MI 48824, USA
}
\date{Accepted XXX. Received YYY; in original form ZZZ}

\pubyear{2021}

\begin{document}
\label{firstpage}
\pagerange{\pageref{firstpage}--\pageref{lastpage}}
\maketitle

\begin{abstract}
Fast radio burst (FRB) 191001 is localised at the spiral arm of a highly star-forming galaxy with an observed dispersion measure (DM) of 507 $\pccc$. Subtracting the contributions of the intergalactic medium and our Milky Way Galaxy from the total DM, one gets an excess of around 200 $\pccc$, which may have been contributed by the host galaxy of the FRB.  It is found in this work that the position of FRB\,191001 is consistent with the distribution of supernovae (SNe) in the spiral arm of their parent galaxies. If this event is indeed due to an SN explosion, then, from the analysis of the SN contributions to the excess DM, a core-collapse (CC) channel is preferred over a thermonuclear runaway. For the CC explosion, depending on the density of the surrounding medium, the age of the central engine that powers the radio burst is within a couple of years to a few decades. However, the observed rotation measure of FRB\,191001 does not confirm the fact that the radio burst has passed through the remnant of a young SN.   
\end{abstract}
\begin{keywords}
shock waves -- stars: magnetars -- radio continuum: transients
\end{keywords}



\section{Introduction}
\label{sec:intro}
Fast radio bursts (FRBs) are transient millisecond duration bright radio pulses of unknown origins \citep{Lorimer07,Petroff19,Cordes19}. The progenitors of these exotic events have been remained elusive since their discovery in the last decade. The millisecond duration of these bursts suggests that the emissions are likely originated from compact sources. This hypothesis is further strengthened by the discovery of galactic FRB\,200428 in association with the magnetar SGR 1935+2154 \citep{chime20_galacticFRB,bochenek20,Mereghetti20,Li20_Xray_galSGR,ridnaia20} that formed due to the collapse of a massive star \citep{gaenler14}. 

\par
The localisation of FRB sources and the identification of host galaxies are important in understanding their progenitors. Among the localised events, FRB\,191001 is discovered at the spiral arm of a highly star-forming galaxy at a redshift of 0.234 \citep{bhandari20_FRB191001}. The observed dispersion measure (DM) of this FRB is 506.9 $\pccc$ \citep{bhandari20_FRB191001}, which may have contributions from the i) host galaxy of the FRB (${\rm DM_{\rm host}}/(1+z)$, where $z$ is the redshift of the host), ii) intergalactic medium (${\rm DM_{\rm IGM}}$), and iii) our Milky Way Galaxy (${\rm DM_{\rm MW}}$). From the galactic models NE2001 \citep{cordes02} and YMW16 \citep{yao17} the DM contribution from our Galaxy, toward the direction of this FRB, is ${\rm DM_{\rm MW}} = 94$ and 81 $\pccc$ \citep{bhandari20_FRB191001}, respectively. The contribution of the intergalactic medium (IGM) is ${\rm DM_{\rm IGM}} \simeq 203 $ $\pccc$ as estimated from \citet{macquart20_nat}. Likewise the DM redshift relation, which is ${\rm DM_{\rm IGM}} \approx 855 z_{\rm max}$ \citep{zhang18} with $z_{\rm max}$ being the maximum value of the redshift of an FRB, predicts a ${\rm DM_{\rm IGM}} \approx 210$ $\pccc$ for the redshift of 0.234. This implies a contribution of the host galaxy ${\rm DM_{\rm host}} \approx 200 \times (1+0.234) \approx 250$ $\pccc$. The host DM will have contributions from the immediate surroundings of the burst and from other ionised media that are encountered by the pulse in the host galaxy. Since the source of this radio burst resides in the spiral arm of its parent galaxy, it is possible that a significant amount of this excess DM has been contributed by the former.

\par 
In the nuclear region of a starburst galaxy, the star formation rate (${\rm SFR}$) is related to the rate of the core-collapse (CC) supernova (SN) (symbolised as ${\rm R}_{\rm CCSN}$) in that galaxy. For a Salpeter initial mass function, with a presumption that stars with masses between 0.1 $\msun$ and 125 $\msun$ are formed in the galaxy, and those typically between 8 $\msun{}$ and 50 $\msun$ undergo CC explosions, it is easy to show that ${\rm R}_{\rm CCSN} = {\rm SFR} ($\msunyr$) \times 7 \times 10^{-3} ~~ yr^{-1}$ \citep{mattila01}. In the case of FRB 191001, an ${\rm SFR}$ of 11.2 $\msunyr$ \citep{bhandari20_FRB191001} implies that there are around 80 SN remnants in the host galaxy of FRB\,191001 that are younger than 1000 yrs. The location of this burst in the spiral arm, the excess DM and the prevalence of SNe in its parent galaxy give us a unique opportunity to explore the origin of this FRB. In the following section, we compare the position of FRB\,191001 with the distribution of SNe in the spiral arm of their host galaxies. In \$ \ref{sec:DM_RM} the evolution of DM and rotation measure (RM) when a radio wave propagates through an SN remnant are reviewed. The paper is closed with a discussion given in \$ \ref{sec:dis}.

\section{SN distribution in spiral arms}
\label{sec:SN_dist_spiral_arm}
\begin{figure*}
\centering
\includegraphics[width=8.0cm,origin=c]{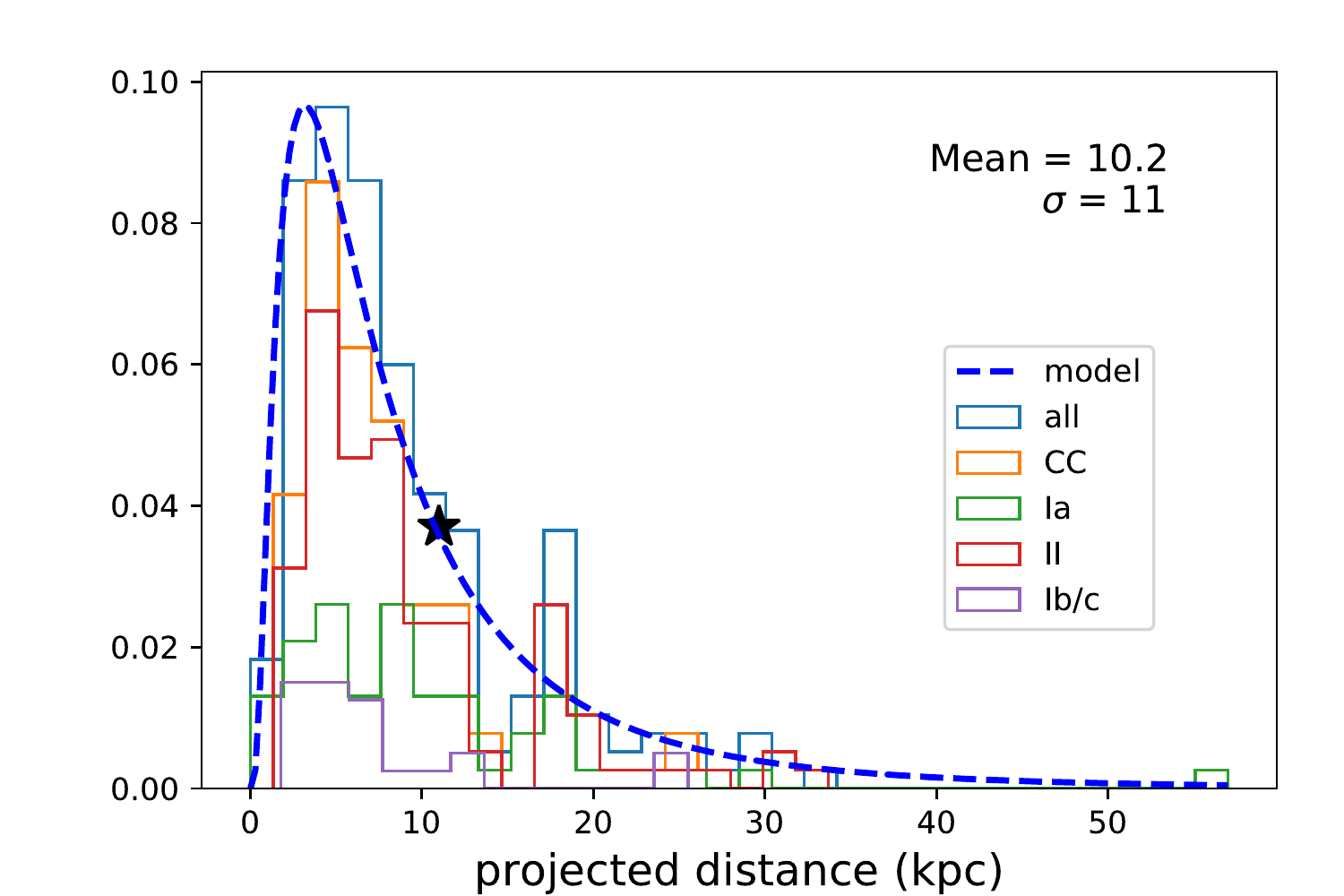}
\includegraphics[width=8.0cm,origin=c]{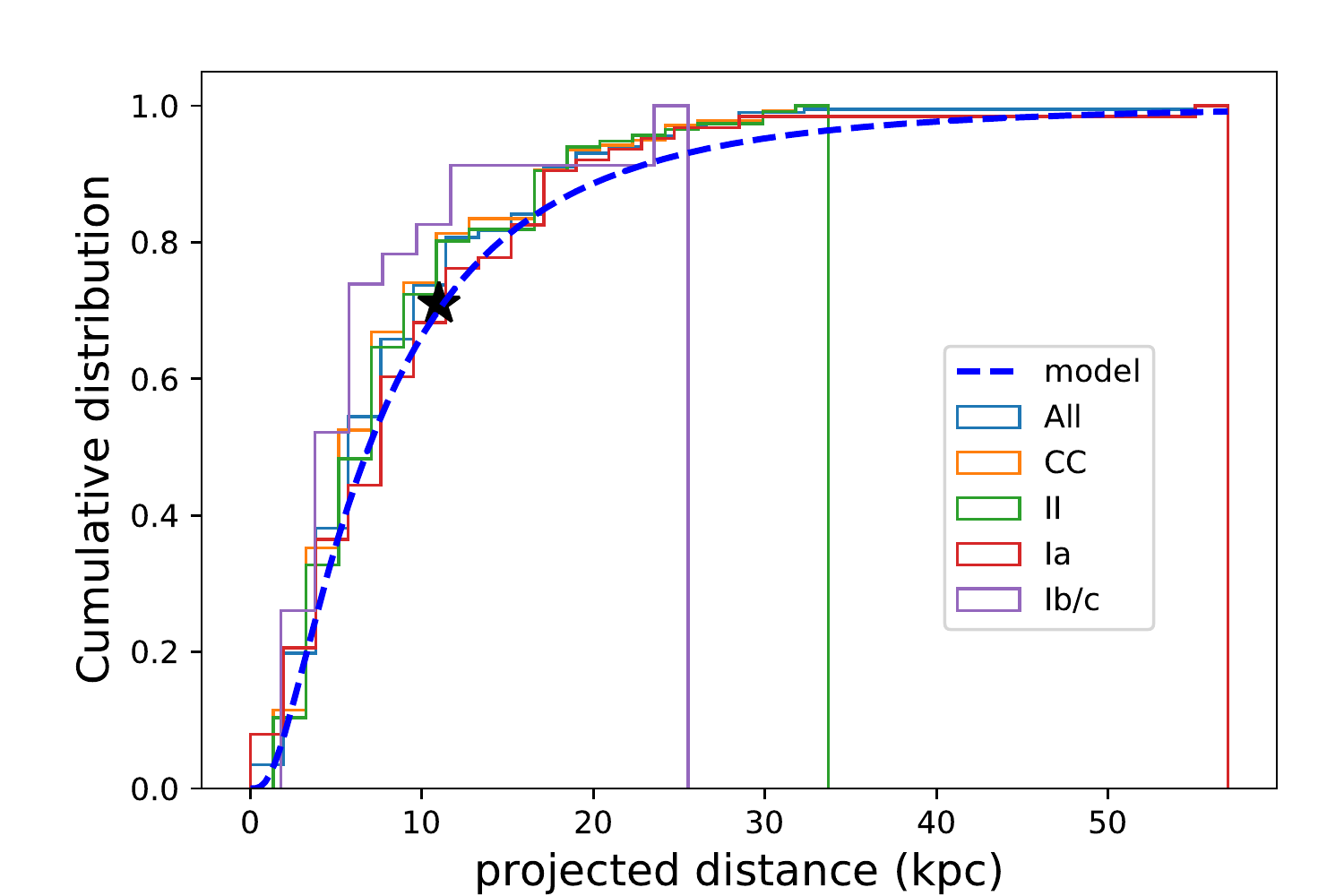}
\caption{Left panel: Normalised distribution of the projected distance of SNe from the centre of their respective host galaxies. The overall distribution is fitted with a log-normal function which is exhibited with a blue dashed curve. The mean and the standard deviation of the distribution are 10.2 kpc and 11 kpc, respectively.  Right panel: Cumulative distributions of different types of SNe along with the log-normal model. The distributions show that around 70\% of SNe has offset $\lsim 11$ kpc. In both panels, the black star represents the position of FRB\,191001, which is at a projected distance of around 11 kpc from the centre of its host galaxy \citep{heintz20}.}
\label{fig_sn_dist}
\end{figure*}

\citet{aramyan16} has given a sample of 215 SNe of different types in the spiral arm of their host galaxies, which we consider for this study. The coordinates of the host galaxies and SNe, and their redshifts are obtained using the SkyCoord class method from Astropy \citep{astropy2013,astropy2018}, from the NASA Extragalactic Database (NED)\footnote{https://ned.ipac.caltech.edu} and bright SN catalog\footnote{https://www.rochesterastronomy.org/snimages/snredshiftall.html}. Among the 215 entries, redshift information for 12 objects is not available in the literature. Besides, due to the large positional uncertainty of an SN event in its host galaxy that object is removed from our study. For the rest of the 202 SNe, the normalised distribution of the projected distance from the centre of their respective host galaxies is shown in the left  panel of Fig.\ref{fig_sn_dist}. Out of these 202 SNe, 139 are CC events while the rest is thermonuclear explosions (Type Ia). Among the 139 CC SNe, 116 are Type II and the rest belongs to Type Ib/c. The different types of SNe are shown with different colours in Fig.\ref{fig_sn_dist}. For the sub-classes, the histograms are plotted such that the sum of the area under CC (II+Ib/c) and Ia is one. The overall distribution is fitted with a log-normal function which is exhibited with a blue dashed curve. The functional form of this function is 
\beq 
f(x) = \frac{1}{x s \sqrt{2\pi}} {\rm exp} \bigg[-\frac{(lnx - \mu)^2}{2s^2}\bigg],
\label{func:log_normal}
\eeq
where $\mu$ and $s^2$ are the expected value and the variance of the variable's natural logarithm. As a result, the mean and the standard deviation of the distribution are calculated as Mean $ = {\rm exp}(\mu + s^2/2)$ and $\sigma = \sqrt{[{\rm exp}(s^2)-1]{\rm exp}(2\mu + s^2)}$. From the fitting, the Mean and $\sigma$ of the SN distribution are estimated to be 10.2 kpc and 11 kpc, respectively. The cumulative distributions of different types of SNe along with the log-normal model are shown in the right panel of Fig.\ref{fig_sn_dist}. In both panels of Fig \ref{fig_sn_dist} the black star represents the position of FRB 191001, which is at a projected distance of around 11 kpc from the centre of its host galaxy \citep{heintz20}. The cumulative distributions demonstrate that around 70\% of SNe has offset $\lsim 11$ kpc. The analysis of this section illustrates that the position of FRB\,191001 is consistent with the distribution of SNe in the spiral arm of their parent galaxies.

\section {Dispersion and Rotation Measures}
\label{sec:DM_RM}
Motivated by the results of \S \ref{sec:SN_dist_spiral_arm}, in this section, we study whether the DM and RM contributions due to CC and Ia events are compatible with the excesses observed for FRB\,191001. When
a source of an FRB is embedded in an SN remnant, the following contributes to the total DM: firstly, the ionised unshocked ejecta; secondly, the shocked shell that contains the shocked ejecta and the shocked circumstellar medium (CSM); and finally, the unshocked ionised CSM. For an SN with a constant density profile ($\rho_{\rm ej}(r)$) for $r < r_{\rm brk}$,  where $r_{\rm brk}$ represents the radius corresponding to the break velocity $v_{\rm brk}$, the DM of the ionised ejecta decreases as $t^{-2}$ for a homologous expansion. In the case of the outer part of the ejecta,  $\rho_{\rm ej}(r) \propto r^{-n}$ , with $n \sim$ 10 \citep{matzner99,kundu17} being the power-law index of the density profile for $r > r_{\rm brk}$. For the shocked shell, the DM is ${\rm DM}_{\rm sh} = \int_{r_{\rm rev}}^{r_s} n_e (r) dr = \int_{r_{\rm rev}}^{r_c} n_e^{\rm rev} (r) dr + \int_{r_{c}}^{r_s} n_e^{s} (r) dr$, where $n_e^{\rm rev} (r)$ and $n_e^{s} (r)$ represent the electron density of the shocked ejecta and the shocked CSM, respectively. $r_s$, $r_{\rm rev}$ and $r_c$ are the radii of forward shock, reverse shock and contact discontinuity, respectively. The total contribution due to the shocked shell in the free-expansion (FE) phase is
\beq 
{\rm DM}_{\rm sh}^{\rm FE} = \frac{4 ~ A~ (\phi_{\rm csm} + \phi_{\rm ej})}{\mu~ m_p (1-s)} ~ D^{\frac{1-s}{n-s}}  ~ t^\frac{(n-3)(1-s)}{(n-s)}
\label{eq_DM_sh_2}
\eeq
(for details see \citet{kundu20}), where the SN interacts with a wind-like or a constant density ambient medium. The density of the CSM can be written as $\rho_{\rm csm} = A r^{-s}$, with $s =$ 2 (0) and $A = \mdot [4\pi v_w]^{-1}$ $(\mu m_p n_{\rm {ISM}})$ for a wind-like (constant density) medium. Here, $\mdot$ and $v_w$ represent the mass-loss rate from the pre-SN star and wind speed at which matter was ejected from the system. $\mu$ and $n_{\rm {ISM}}$ are the mean atomic weight and particle density of the ambient medium, respectively. $m_p$ represents the mass of a proton. $D = \xi v_{\rm brk}^n/A$ with $\xi$ being a constant. For $\beta = r_{\rm rev}/r_c$ and $\alpha = r_s/r_c$, $\phi_{\rm csm} = (\alpha^{1-s} -1 )$ and $\phi_{\rm ej} = ~\frac{(n-3)(n-4)}{(3-s)(4-s)} ~ (1- \beta^{1-s} )$. 
At the end of the FE, the SN evolves into the Sedov-Taylor (ST) phase. If $T_{\rm{FE}}$ represents the duration of the FE phase then the evolution of the DM in the ST phase is written as 
\beq
{\rm DM}_{\rm sh}^{\rm ST} = \frac{4 ~ A~ (\phi_{\rm csm} + \phi_{\rm ej})}{\mu~ m_p (1-s)} ~ D^{\frac{1-s}{n-s}} ~ T_{\rm FE}^{\frac{(n-5)(3-s)(1-s)}{(n-s)(5-s)}} ~ t^{\frac{2(1-s)}{5-s}}.
\label{eq:DM_ST_fw_rev}
\eeq
As shocks amplify magnetic fields \citep{bykov13,caprioli14}, the RM due to the shell, in FE and ST phases are 
\beq
{\rm RM}_{\rm sh}^{\rm FE}  =\Omega \frac{2 \alpha  (n-3)}{(2-3s)(n-s)} t^{\frac{2(n-6)+s(11-3n)}{2(n-s)}},
\label{eq:RM_FE_fw_rev}
\eeq
and
\beq 
{\rm RM}_{\rm sh}^{\rm ST}  = \Omega \frac{4 \alpha}{(5-s)(2-3s)} T_{\rm FE}^{\frac{(n-5)(3-s)(4-3s)}{2(n-s)(5-s)}} ~ t^{\frac{2s+1}{s-5}},
\label{eq:RM_ST_fw_rev}
\eeq
respectively, \citep{kundu20}, where $\Omega = 4 \frac{e^3}{2 \pi m_e^2 c^4} \frac{({9 \pi \epsilon_{\rm B} A^3})^{1/2}}{\mu m_p} (\psi_{\rm csm} + \psi_{\rm ej}) D^{\frac{(4-3s)}{2(n-s)}} $, $\epsilon_{\rm B}$ represents the fraction of post shock energy that goes to magnetic fields, $ \psi_{\rm csm} =[ \alpha^{(1-3s/2)} -1]$ and  $ \psi_{\rm ej} = \frac{(n-3)(n-4)}{(3-s)(4-s)} ~ (1 - \beta^{(1-3s/2)})$. $c$, $m_e$ and $e$ represent the velocity of light in vacuum, electron mass and charge, respectively.

\par 
After the explosion of a star, radiation that comes from the shock breakout, shocked shell, and hot ejecta ionises a fraction of the CSM. In the case of a CC scenario, the DM contribution from the ambient medium is
\beq 
\rm DM_{\rm CSM} = \frac{A}{r_s} \propto t^{-(n-3)/(n-2)}. 
\label{eq_DM_csm}
\eeq
While for a constant density medium ($s = 0$) 
\beq 
\rm DM_{CSM} = \mu ~ \Big(\frac{n_{\rm ISM}}{1 \ccc}\Big) ~ \Big(\frac{\Delta r}{1 {\rm pc}}\Big) ~ pc ~ \ccc,
\label{eq_DM_csm_2}
\eeq
\citep{kundu20}, where $\Delta r = r_{max} - r_s$ with $r_{max}$ being the radius up to which the medium is ionised, and $\mu \simeq$ 1 for a medium that contains hydrogen and helium with solar abundances. Nonetheless, this media would have an almost null contribution to the total RM as the magnetic field in the CSM is not expected to be oriented along a given direction.

\subsection{Core-collapse explosion }
\label{subsec_cc}
\begin{figure*}
\centering
\includegraphics[width=8.0cm,origin=c]{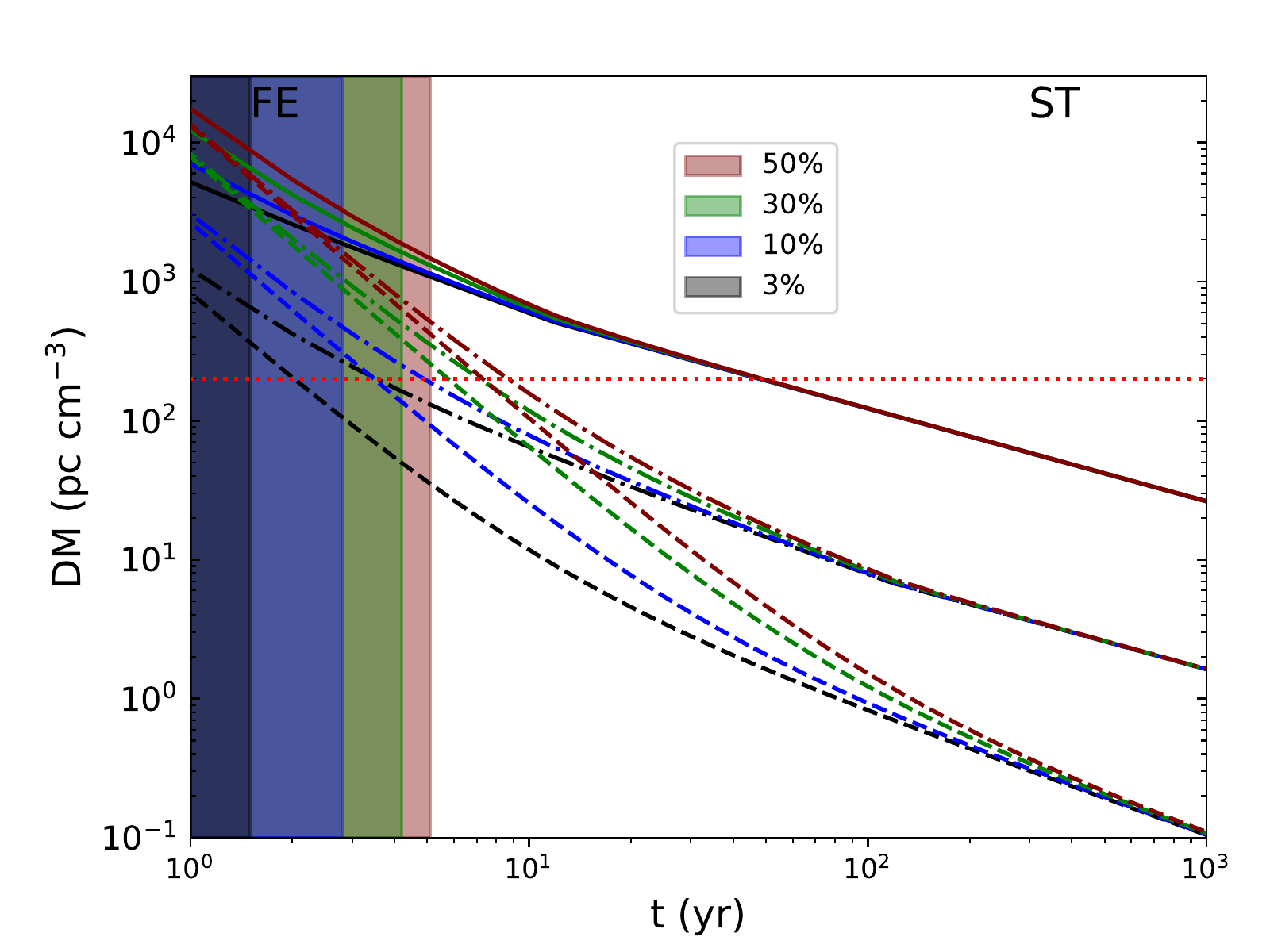}
\includegraphics[width=8.0cm,origin=c]{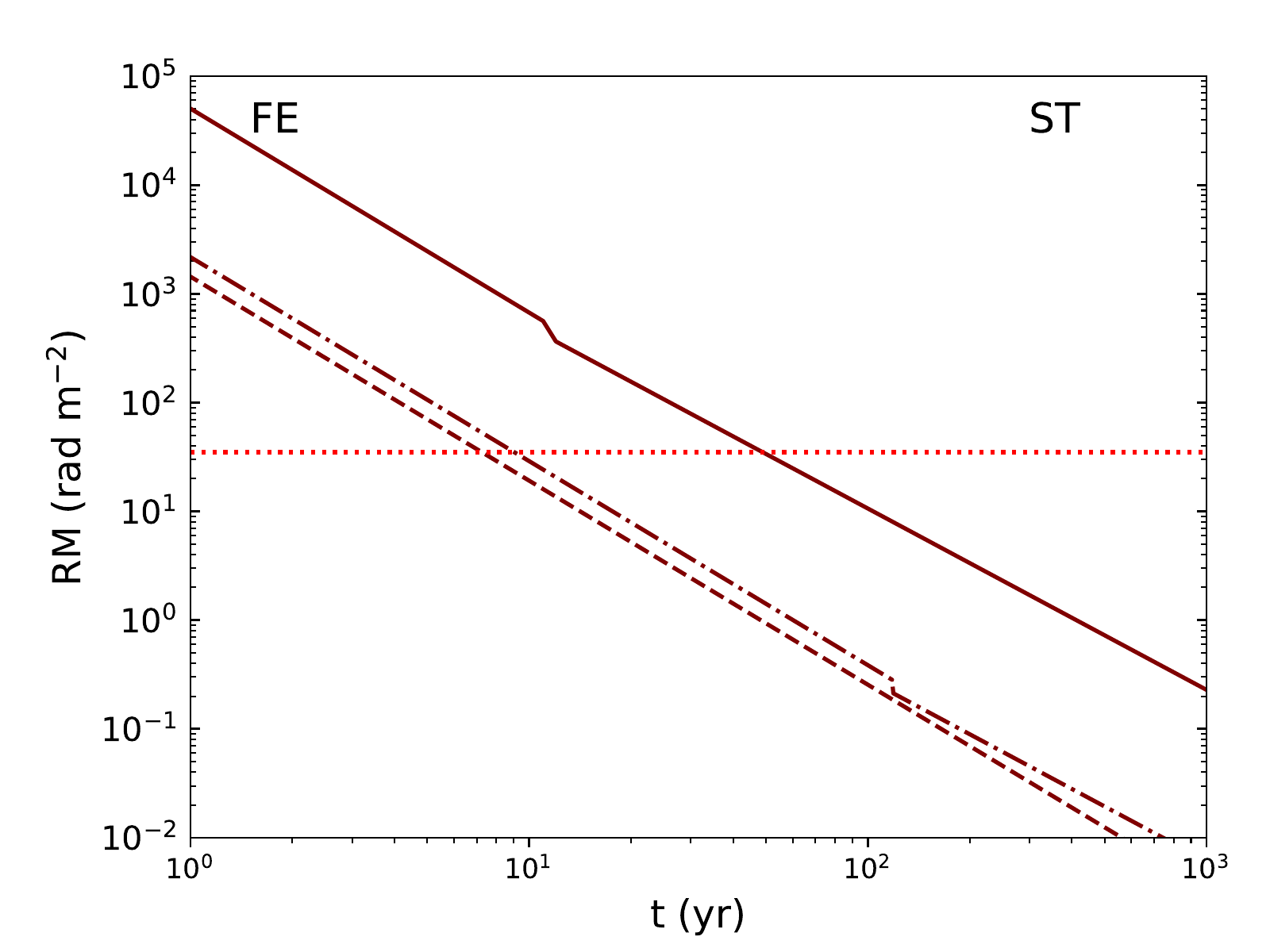}
\caption{Evolution of the DM (left panel) and RM (right panel) as a function of time for the CC scenario. The solid, dash-dotted and dashed lines demonstrate the evolution when the SN ploughs through an ambient medium characterised by a $\mdot = 1\times 10^{-4}$ \msunyr,   $1\times 10^{-5}$  \msunyr and $1\times 10^{-6}$  \msunyr, respectively, for a wind velocity $v_w $ of 10 \kms. Left panel: The maroon, green, blue and black lines exhibit the evolution when the ejecta is ionised by 50$\%$, 30$\%$, 10$\%$ and 3$\%$, respectively. The shaded regions show the time until the medium remains optically thick to a 1 GHz signal. The red dotted horizontal line represents a DM value of 200 $\pccc$, which is the excess DM of FRB\,191001 (see \$ \ref{sec:dis}). Right panel: RM for the three mass-loss rates when 50\% of the ejecta is ionised. The red dotted horizontal line represents a RM value of 35 rad m$^{-2}$, which is the inferred excess RM of FRB\,191001 (see \$ \ref{sec:dis}).}
\label{fig_CC_wind}
\end{figure*}

Stars more massive than 8 -10 $\msun$ usually end their life in CC explosions. The pre-SN star loses matter mainly through strong winds. In this work, we consider a CC explosion with ejecta of around 5 $\msun$ with $v_{\rm brk} \sim 5000\,\kms$ and explosion energy $E_{\rm k} = 10^{51}$\,erg. 
 For a significant fraction of CC events, mass-loss rates vary in the range $10^{-4} - 10^{-6}$ \msunyr, while the typical wind velocity is $\sim$ 10 \kms \citep{smith14}. With a power-law index $n = 10$ for the outer part of the ejecta, the evolution of the  $\rm DM$ (a sum of the contribution from the ionised ejecta, shocked shell and unshocked CSM) and $\rm RM$ (contribution from the shocked shell) are shown in Fig.\ref{fig_CC_wind}. The solid, dash-dotted, and dashed lines demonstrate the evolution when the SN ploughs through an ambient medium characterised by a $\mdot = 1\times 10^{-4}$ \msunyr,   $1\times 10^{-5}$  \msunyr and $1\times 10^{-6}$  \msunyr, respectively, for a wind velocity $v_w $ of 10 \kms. The maroon, green, blue and black lines exhibit the evolution when the ejecta is ionised by 50$\%$, 30$\%$, 10$\%$ and 3$\%$, respectively. In the case of CC SN\,1993J \citet{chevalier16} found that the ejecta is ionised by around 3$\%$. However, to examine the effect of a higher ionisation fraction on the evolution of the DM we consider cases where the unshocked inner part is ionised from 3$\%$ to 50$\%$.            

\par 
The shocks are initially in the FE phase. The duration of this phase can be calculated as $T_{\rm{FE}} = \Lambda^\frac{1}{(3-s)} \beta^\frac{n-s}{3-s} v_{\rm brk}^\frac{s}{3-s}$, where $\Lambda = \frac{4 \pi}{\theta}\left(\frac{3-s}{n-3} \right)\frac{\rho_{ej,in}}{\Delta} \frac{\beta^{3-n}}{\alpha^{3-s}} r_{\rm o,w}^{2-s}$ \citep{kundu17}. Here $\Delta = \mdot/v_w$ ($n_{\rm ISM}$) for a wind-like  (constant density) medium, $\theta$ represents a ratio of the swept-up ejecta and the swept-up ambient mass. $r_{\rm o,w}$ is a reference radius, which is assumed to be $2.5 \times 10^{14}$ cm ($10^{15}$ cm) for CC (Ia) events, and $\rho_{\rm ej,in}$ represents the density of the inner ejecta. The value of $T_{\rm{FE}}$ for $\mdot = 1\times 10^{-4}$ \msunyr, $1\times 10^{-5}$  \msunyr and $1\times 10^{-6}$  \msunyr, are around 12, 119 and 1200 yrs, respectively, for $v_w = 10$ \kms. The evolution of the RM for the three mass-loss rates is accordingly shown in the right panel of Fig.\ref{fig_CC_wind} when around 50\% of the ejecta is ionised. For these $\mdot$, the $\epsilon_{\rm B}$ values are in the range $10^{-8}$ to $10^{-11}$.

\subsection{Thermonuclear (Type Ia/ SN Ia/ Ia) explosion }
\label{subsec_Ia}
In general, SNe Ia are considered to be the explosion of carbon-oxygen white dwarfs (WDs) \citep{hoyle60}. There are two main progenitor channels, namely the single degenerate (SD) and the double degenerate (DD), that are thought to lead to a Type Ia. 
In the former scenario, the WD accretes matter from a non-degenerate companion, which triggers an explosion when the WD reaches close to the Chandrasekhar (CH) mass limit \citep{whelan73}. While, the second channel, called the DD, consists of two spiralling sub-CH WDs that eventually merge and lead to an Ia explosion under proper physical conditions \citep{iben84,webbink84}. In this work, the density profile of the Ia ejecta is considered similar to that given by the N100 \citep{ropke12,sei13} and violent merger models of \citet{pakmor12}. Interestingly, these density structures have a flat inner part which decreases sharply as a power-law beyond a break velocity $v_{\rm brk}$. For both SD and DD channels, we consider ejecta of around 1 $\msun$ with $E_{\rm k} = 10^{51}$\,erg, $v_{\rm brk} \sim 10000\,\kms$ and $n = 10$.

\par
\begin{figure*}
\centering
\includegraphics[width=8.0cm,origin=c]{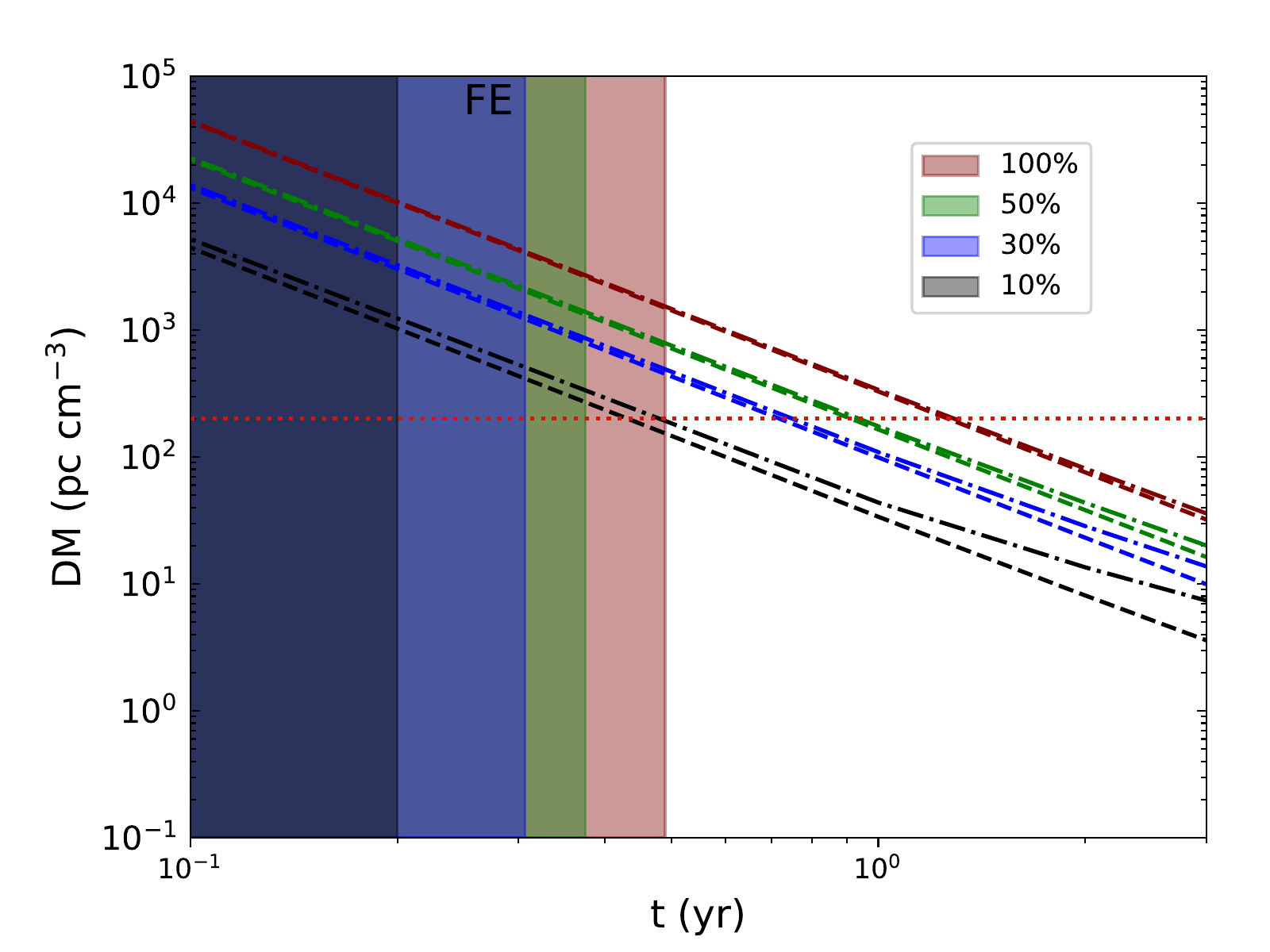}
\includegraphics[width=8.0cm,origin=c]{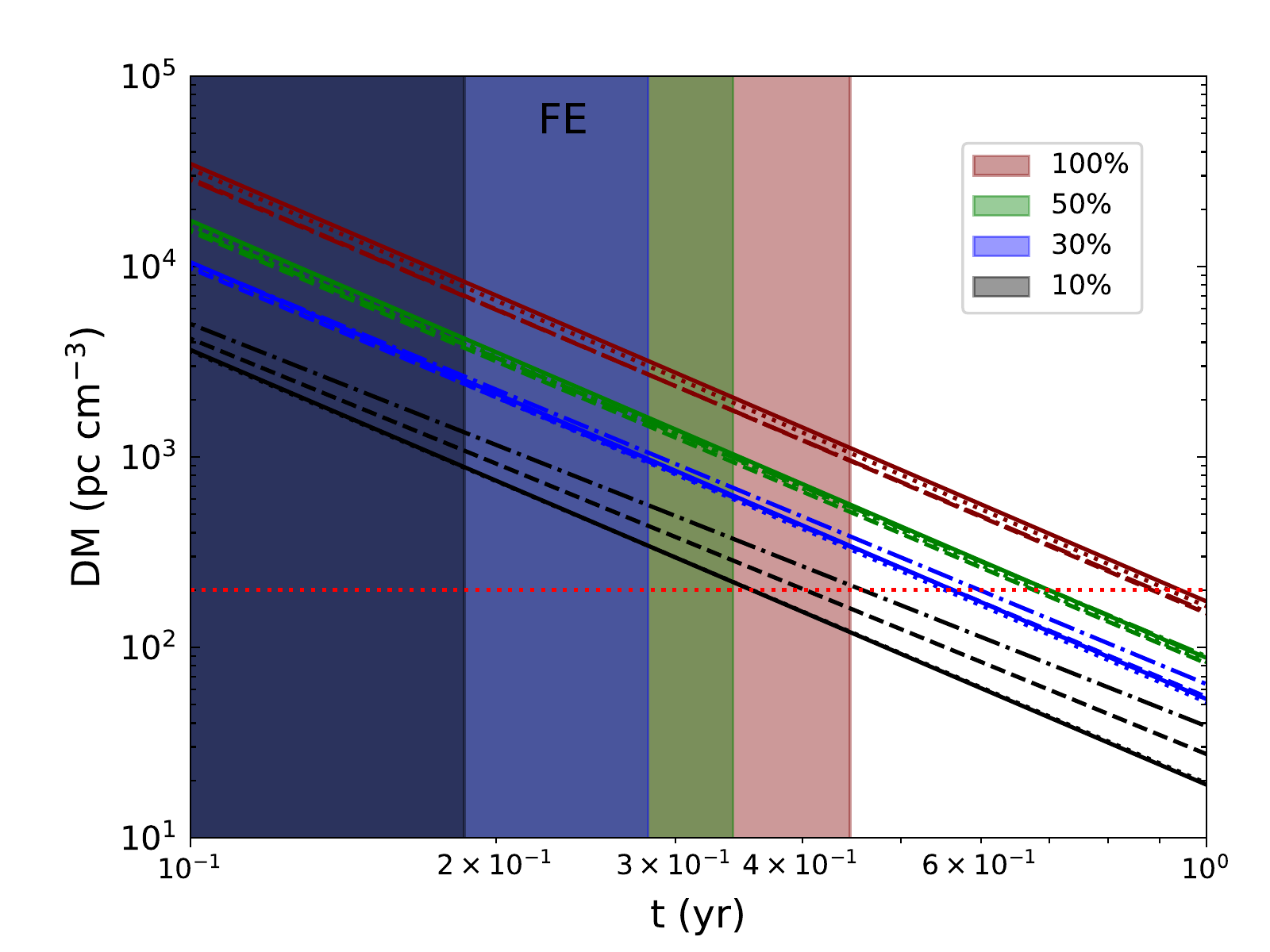}
\caption{DM evolution for the SD (left panel) and DD (right panel) channels. The red dotted horizontal line represents a DM value of 200 $\pccc$, which is the excess DM of FRB\,191001 (see \$ \ref{sec:dis}). Left panel: The dash-dotted and dashed lines demonstrate the evolution when the SN ploughs through an ambient medium characterised by a $\mdot = 1\times 10^{-5}$ \msunyr and   $1\times 10^{-6}$  \msunyr, respectively, for a wind velocity $v_w $ of 100 \kms. Right panel: the dash-dotted, dashed, dotted (excluding the horizontal red line) and solid lines represent the cases when the SN interacts with an ambient medium having a $n_{\rm ISM} =$1000 \ccc, 500 \ccc, 100 \ccc and 50 \ccc, respectively. The maroon, green, blue and black lines exhibit the evolution when the ejecta is ionised by 100$\%$, 50$\%$, 30$\%$ and 10$\%$, respectively. In both panels the shaded regions show the time until the medium remain optically thick to a 1 GHz signal. }
\label{fig_Ia_wind_ISM}
\end{figure*}

While in the case of the SD scenario, mass loss from the non-degenerate companion, either in the form of strong wind or due to Roche-lobe overflow creates a wind-like CSM around the pre-SN system, for DD channel a constant density ambient medium, which is a characteristic of the interstellar medium, exists. 
The Radio and X-ray observations of Type Ia exhibit that for a significant number of thermonuclear events the $\mdot$ is around $10^{-6}$ \msunyr for a presumed $v_w =100$ \kms, and the $n_{\rm ISM}$ is in the range of 50 to 1000 \citep{chomiuk16,margutti12}. In the left panel of Fig.\ref{fig_Ia_wind_ISM}, the evolution of the DM is shown for the SD scenario where the dash-dotted, and dashed lines demonstrate the evolution when the SN ploughs through an ambient medium characterised by a $\mdot = 1\times 10^{-5}$ \msunyr and   $1\times 10^{-6}$  \msunyr, respectively, for $v_w $ of 100 \kms. The maroon, green, blue and black lines exhibit the cases when the ejecta is ionised by 100$\%$, 50$\%$, 30$\%$ and 10$\%$, respectively. For $\mdot = 1\times 10^{-5}$ \msunyr and $1\times 10^{-6}$  \msunyr the shocks will be in the FE phase for around 120 and 1200 yrs after the explosion. In the case of the DD scenario, the DM as a function of time is demonstrated in the right panel of Fig.\ref{fig_Ia_wind_ISM}, where the dash-dotted, dashed, dotted (excluding the horizontal one) and solid lines represent the cases when the SN interacts with an ambient medium with $n_{\rm ISM} =$1000 \ccc, 500 \ccc, 100 \ccc and 50 \ccc, respectively. For these ISM densities, the SN will be in the FE phase around 11 yr, 14 yr, 24 yr, 30 yr, respectively.

\section{Discussion} 
\label{sec:dis}
Besides the immediate surroundings of an FRB source, the host galaxy of the burst also contributes to ${\rm DM_{\rm host}}$. For non-repeating FRBs, \citet{zhang_g_q_20} infer a host galaxy contribution of 30 - 70 $\pccc$ in the redshift range 0.1 to 1.5. The position of the one-off FRB\,191001 is at the spiral arm of its host galaxy. Hence, we assume that the DM contribution from its parent galaxy is around 50 $\pccc$. This leads to an excess DM of about 200 $\pccc$. 
The reason for this excess DM could be a much higher contribution from the immediate surroundings of the FRB, similar to that is inferred as the source of excess DM for FRB\,190608 \citep{chittidi20} and FRB\,121102 \citep{tendulkar17}. It may also be possible that the FRB has traversed through an unusual dense path along its way to us or a combination of both are the source of this excess DM. In this letter, the first scenario is examined in detail.
Studies of host galaxies of different types of SNe demonstrate that star-forming spiral galaxies host a large number of CC SNe \citep{bergh05,hakobyan14}. With an ${\rm SFR} \approx 11$ $\msunyr$, it is expected that the parent galaxy of FRB\,191001 has a number of CC events. Interestingly, the position of FRB\,191001 is found to be consistent with the SN distribution in the spiral arm of the host galaxies as demonstrated in Section \ref{sec:SN_dist_spiral_arm}, and Fig.\ref{fig_sn_dist}. 
Motivated by these facts, we study the possibility of FRB\,191001 being the results of an SN event based on its excess DM, RM and other observed properties.

\par
When a star explodes, initially, the ionised unshocked ejecta dominate the evolution of the DM 
(see \citet{kundu20} for details). As the density of the ionised particles is high in the beginning, this medium remains opaque to a GHz signal at this phase due to the free-free absorption. The free-free absorption coefficient, at a frequency $\nu$, for a plasma with an electron density $n_e$ and temperature $T$ can be written as 
$\alpha_{\nu}^{\rm ff} = \frac{4 e^6}{3 c m_e k_B } \bigg(\frac{2 \pi}{3 k_B m_e}\bigg)^{1/2} T^{-3/2} Z^2  n_e  n_i  \nu^{-2}  \bar{g}_{\rm ff}$
\citep{ribicki79}, where $\bar{g}_{\rm ff}$ represent the velocity average gaunt factor. $Z$ and $n_i$ are the atomic number and density of the ion in that medium, respectively. $k_B$ represents the Boltzmann constant. In the left panel of Fig.\ref{fig_CC_wind}, the black, blue, green and maroon shaded regions show the time until the medium remains optically thick to a 1 GHz signal when the ejecta is ionised by 3$\%$, 10$\%$, 30$\%$ and 50$\%$, respectively. For Ia events, in Fig. \ref{fig_Ia_wind_ISM}, the same shaded regions represent the cases when the ejecta is ionised by 10$\%$, 30$\%$, 50$\%$ and 100$\%$, respectively. In these figures, the red dotted horizontal line represents a DM value of 200 $\pccc$, which is the excess DM of FRB\,191001. Unlike CC events central remnants are not usually formed in the case of SNe Ia. Therefore, if an Ia generates radio bursts at the time of the explosion, it is expected that the radio waves would be detected by the latest when the SN becomes optically thin to the GHz signal. In the case of FRB\,191001, the evolution of the DM for both SD and DD scenarios shows that the DM contribution is around 1000 $\pccc$ when the free-free optical depth is $\sim$ 1 (see Fig. \ref{fig_Ia_wind_ISM}), for any value of ejecta ionisation. This  is almost double the total observed DM of this FRB. It, therefore, seems unlikely that FRB\,191001 was the result of a thermonuclear explosion. Besides, no SNe Ia, with ages in the range 0 to $\sim$ 1 yr, are reported around the location of FRB\, 191001 (RA 21h:33m:24.373s, Dec -54$^{\circ}$:44$\arcmin$:51.86$\arcsec$) by the SN surveys effective in this part of the sky, e.g., the Gaia \citep{gaia16} and ASAS-SN \citep{shappee14,kochanek17}. However, it should be noted that, with a redshift of 0.234 there are fair chances that an Ia event is missed by these surveys as, for SNe, these surveys are typically sensitive to redshifts $< 1.5$ \citep{belokurov03}.

\par 
For the CC scenario, the excess DM of $\sim 200$ $\pccc$ could be contributed by an SN remnant around 50 yr after the explosion when $\mdot$ is $1\times 10^{-4}$ \msunyr (see left panel of Fig.\ref{fig_CC_wind}). This amount of DM is contributed by a comparatively young SN with an age of $\sim $ 2 yr to 10 yr when the density of the surrounding medium is characterised by $\mdot=1\times 10^{-5}$ and $1\times 10^{-6}$ \msunyr and the ionisation fraction of the ejecta varies in the range 3\% to 50\%. 
The RM of FRB\,191001 is reported to be 55.5 rad m$^{-2}$ \citep{bhandari20_FRB191001}. Since the line of sight of this FRB does not pass through our Galactic plane it is assumed that around 20 rad m$^{-2}$ is contributed by our Milky Way Galaxy. Therefore, the remnant contributes around 35 rad m$^{-2}$ on the presumption that almost nothing is contributed by the IGM and the parent galaxy. The red dotted horizontal line in the right panel of Fig.\ref{fig_CC_wind} represents the remnant's contribution. With the constraints on the age of the SN from DM evolution, it is found that to have a contribution as small as 35 rad m$^{-2}$, the $\epsilon_{\rm B}$ of the shock should be in the range $10^{-8}$ to $10^{-11}$ when 50\% of the ejecta is ionised. This is many orders of magnitude lower compared to those that are usually seen in SN radio shells. However, it should be noted that the measured RM may vary significantly from its absolute value, as the resultant RM depends on the projection of magnetic fields along the line of sight. Besides, for a line of sight that is almost perpendicular to the magnetic field of the shell, the $\epsilon_{\rm B}$ is expected to be small. As a result, the RM can not confirm the presence of an SN remnant around the FRB source. However, it does not rule out a CC scenario either. It is, therefore, possible that FRB 191001 has originated from a central engine that formed as a result of a CC explosion.

\section*{Acknowledgements}
 I thank R. Bhat, C. James, S. McSweeney, N. Swainston and K. Smith for useful discussions. I acknowledge the Australian Research Council (ARC) grant DP180100857. 

\section*{Data Availability}
The data underlying this article will be shared on reasonable request to the corresponding author.

\bibliographystyle{mnras}
\bibliography{referns} 

\bsp	
\label{lastpage}
\end{document}